\documentclass[conference]{IEEEtran}
\IEEEoverridecommandlockouts

\usepackage{float}

\title{Liquid Crystal-Based RIS Loss-Trade-Off Analysis}

\author{
\IEEEauthorblockN{Bowu Wang, Mohamadreza Delbari, Robin Neuder, Alejandro Jim\'{e}nez-S\'{a}ez, and Vahid Jamali}
\IEEEauthorblockA{
Technical University of Darmstadt (TUD), Darmstadt, Germany.
\vspace{-0.5cm}
}
\thanks{Wang, Delbari and Jamali’s work was supported in part by the Deutsche Forschungsgemeinschaft (DFG, German Research Foundation) within the Collaborative Research Center MAKI (SFB 1053, Project-ID 210487104) and in part by the LOEWE initiative (Hesse, Germany) within the emergenCITY Centre under Grant LOEWE/1/12/519/03/05.001(0016)/72. Neuder and Jim\'{e}nez-S\'{a}ez's work was supported by the Deutsche Forschungsgemeinschaft (DFG, German Research Foundation) – Project-ID 287022738 – TRR 196 MARIE within project C09.}
}

\usepackage{amsfonts,amsmath,amsthm,bm}
\usepackage{cite}
\usepackage{amsmath,amssymb,amsfonts}
\usepackage{algorithmic}
\usepackage{graphicx}
\usepackage{textcomp}
\usepackage{xcolor}
\usepackage[T1]{fontenc}
\usepackage{comment}
\usepackage{eucal}
\usepackage{algorithm}
\usepackage{algorithmic}
\usepackage[acronyms,nonumberlist,nopostdot,nomain,nogroupskip]{glossaries}
\usepackage[font=footnotesize]{caption}
\usepackage{subcaption}
\usepackage{booktabs,flushend}
\usepackage{xcolor}
\definecolor{AlmostWhite}{gray}{0.5}

\newcommand{\defeq}{\triangleq}

\newcommand{\e}{\mathsf{e}}
\newcommand{\jj}{\mathsf{j}}
\newcommand{\Herm}{\mathsf{H}}
\newcommand{\Trans}{\mathsf{T}}

\newcommand{\bA}{\mathbf{A}}
\newcommand{\bx}{\mathbf{x}}

\newcommand{\bs}{\mathbf{s}}
\newcommand{\bS}{\mathbf{S}}

\newcommand{\bH}{\mathbf{H}}
\newcommand{\ba}{\mathbf{a}}

\newcommand{\bu}{\mathbf{u}}

\newcommand{\bh}{\mathbf{h}}

\newcommand{\bq}{\mathbf{q}}
\newcommand{\bp}{\mathbf{p}}

\newcommand{\kk}{\kappa}

\newcommand{\bSigma}{\boldsymbol{\Sigma}}
\newcommand{\bGamma}{\boldsymbol{\Gamma}}
\newcommand{\bmu}{\boldsymbol{\mu}}

\newcommand{\blambda}{\boldsymbol{\lambda}}

\newcommand{\Ex}{\mathbb{E}}

\newcommand{\diag}{\mathrm{diag}}
\newcommand{\real}{\mathrm{Re}}
\newcommand{\imag}{\mathrm{Im}}
\newcommand{\tr}{\mathrm{tr}}
\newcommand{\rank}{\mathrm{rank}}

\newcommand{\tx}{\mathrm{tx}}
\newcommand{\thr}{\mathrm{thr}}
\newcommand{\SNR}{\mathrm{SNR}}

\newcommand{\RIS}{\mathrm{RIS}}

\newcommand{\dB}{\mathrm{dB}}

\def\bomega{\boldsymbol{\omega}}

\def\bzero{\boldsymbol{0}}
\def\bone{\boldsymbol{1}}
\def\Cset{\mathbb{C}}

\def\Rset{\mathbb{R}}

\def\LOS{\mathrm{LOS}}

\def\tmax{\mathrm{max}}
\def\tx{\mathrm{tx}}
\def\rx{\mathrm{rx}}
\def\BS{\mathrm{BS}}
\def\RIS{\mathrm{RIS}}

\def\SNR{\mathrm{SNR}}
\def\eff{\mathrm{eff}}

\def\FoM{\mathrm{FoM}}

\def\thr{\mathrm{thr}}

\def\sCN{\mathcal{CN}}

\def\Wset{\mathcal{W}}

\def\Pset{\mathcal{P}}
\def\bigO{\mathcal{O}}
\usepackage{xcolor}

\newacronym{RIS}{RIS}{reconfigurable intelligent surface}
\newacronym{QoS}{QoS}{quality of service}
\newacronym{LC}{LC}{liquid crystal}
\newacronym{SNR}{SNR}{signal to noise ratio}
\newacronym{TDMA}{TDMA}{time-division multiple-access}
\newacronym{BS}{BS}{base station}
\newacronym{MU}{MU}{mobile user}
\newacronym{ME}{ME}{mobile eavesdropper}
\newacronym{NF}{NF}{near-field}
\newacronym{Tx}{Tx}{transmitter}
\newacronym{Rx}{Rx}{receiver}
\newacronym{AWGN}{AWGN}{additive white Gaussian noise}
\newacronym{w.r.t.}{w.r.t.}{with respect to}
\newacronym{RDE}{RDE}{Reaction-Diffusion Equation}
\newacronym{PDE}{PDE}{partial differential equation}
\newacronym{UPA}{UPA}{uniform planar array}
\newacronym{AO}{AO}{alternative optimization}
\newacronym{SOCP}{SOCP}{second-order cone programming}
\newacronym{AoD}{AoD}{angle of departure}
\newacronym{LOS}{LOS}{line of sight}
\newacronym{nLOS}{nLOS}{non-LOS}
\newacronym{MIMO}{MIMO}{multiple-input multiple-output}
\newacronym{RS}{SR}{secure rate}
\newacronym{SDP}{SDP}{semi-definite programming}
\newacronym{6G}{6G}{sixth generation}
\newacronym{CSI}{CSI}{channel state information}

\newacronym{PIN}{PIN}{positive-intrinsic-negative}
\newacronym{RF}{RF}{radio frequency}
\newacronym{MEMS}{MEMS}{micro-electro-mechanical system}
\newacronym{mmWave}{mmWave}{millimeter wave}
\newacronym{FOM}{FoM}{figure of merit}
\newacronym{MRT}{MRT}{maximum ratio transmission}

\begin{document}

\maketitle

\begin{abstract}
\Gls{LC} technology has emerged as a promising solution for large \glspl{RIS} at \gls{mmWave} bands, offering advantages such as low power consumption, scalability, and continuously tunable phase shifts. For \gls{LC}-\gls{RIS} based on the delay-line architecture, i.e., with dedicated phase shifters, there exists a trade-off between the maximum achievable phase-shift range and the corresponding insertion loss, which has not been studied for \gls{LC}-\gls{RIS}-assisted wireless systems yet.
In this paper, we investigate this trade-off where a \gls{BS} and an \gls{RIS} are configured to minimize the transmit power while satisfying a given \gls{QoS} for a number of users. Simulation results reveal a fundamental trade-off between the total transmit power and the achievable data rate as a function of the \gls{LC} phase-shift range.

\end{abstract}
\glsresetall
\section{Introduction}

\Glspl{RIS} have emerged as a promising technology to enhance wireless communication performance by enabling programmable and energy-efficient manipulation of electromagnetic waves. \Gls{LC}-based implementations of RIS offer significant advantages, such as low power consumption, cost efficiency, and the ability to achieve continuously tunable phase shifts particularly for high frequency bands \cite{jimenez2023reconfigurable}. However, there is a trade-off for delay-line architecture LC-based phase shifters between the maximum achievable phase-shift range and the associated insertion loss \cite{jakoby2020microwave,neuder2024architecture}. In particular, shorter phase shifter lengths reduce insertion loss but also constrain the maximum achievable phase shift, which can fall below the full \(2\pi\) range and negatively impact the performance. 

Several studies in the literature have investigated the energy efficiency of \gls{RIS}-assisted wireless communications in general. For example, \cite{Huang2019} introduced a framework for optimizing the energy efficiency of \gls{RIS}-assisted networks, showing that the joint optimization of transmit power and \gls{RIS} phase shifts is crucial for mitigating hardware-related power consumption. This theoretical foundation was later confirmed in experimental evidence by \cite{Tang2021}, who conducted extensive measurements and developed path loss models that empirically confirmed the impact of real-world reflection losses. More recent analyses, such as that by \cite{Singh2023}, have advanced this line of inquiry by explicitly modeling performance degradation caused by phase-dependent insertion loss. To address these intrinsic losses in passive surfaces, alternative architectures have been proposed; for instance, \cite{Ndjiongue2021} investigated active, power-amplifying \gls{RIS} designs that offset losses at the expense of increased on-device power consumption, thereby revealing another dimension of the same fundamental trade-off.


To the best of the authors' knowledge, this trade-off has not been investigated in \gls{LC}-\gls{RIS}-assisted wireless communications yet. In this work, we investigate the loss performance trade-off in LC-RIS-assisted wireless communication systems. We formulate a power minimization algorithm that dynamically adjusts the RIS phase shifts while accounting for the insertion loss and restricted phase-shift range. Simulation results demonstrate that there exists an LC phase shifter length that achieves the scenario-dependent optimal trade-off between the total transmission power and the data rate, thus enabling energy-efficient wireless communications with LC-RIS.

\textit{Notation:} Bold capital and small letters are used to denote matrices and vectors, respectively.  $(\cdot)^\Trans$, $(\cdot)^\Herm$, $\rank(\cdot)$, and $\tr(\cdot)$ denote the transpose, Hermitian, rank, and trace of a matrix, respectively.  Moreover, $\diag(\bA)$ is a vector that contains the main diagonal entries of matrix $\bA$. $\|\bA\|_*=\sum_i \sigma_i$, $\|\bA\|_2=\max_i \sigma_i$, $\|\bA\|_F$, and $\blambda_{\max}(\bA)$ denote the respectively nuclear, spectral, and Frobenius norms of a Hermitian matrix $\bA$, and eigenvector associated with the maximum eigenvalue of matrix $\bA$, where $\sigma_i,\,\,\forall i$, are the singular values of $\bA$. Furthermore, $[\bA]_{m,n}$ and $[\ba]_{n}$ denote the element in the $m$th row and $n$th column of matrix $\bA$ and the $n$th entry of vector $\ba$, respectively. 
Moreover, $\Rset$ and $\Cset$ represent the sets of real and complex numbers, respectively, and $\jj$ is the imaginary unit. 
$\Ex\{\cdot\}$ is expectation and $\mathcal{CN}(\bmu,\bSigma)$ denotes a complex Gaussian random vector with mean vector $\bmu$ and covariance matrix $\bSigma$. $\mathrm{rand}(N)$ denotes a $N\times1$ vector where each element is generated independently and uniformly from 0 to 1. Finally, $\bigO(\cdot)$ represents the big-O notation and $|\Pset|$ is the cardinality of set $\Pset$.

\section{System, Channel, and Data Rate Models}
\label{system model}
In this section, we first present the system model for the \gls{MU}. We then describe the channel model, followed by the data rate model, which serves as the \gls{MU}’s performance metric.

\subsection{System Model}
We consider a narrow-band downlink system with a \gls{BS} employing \( N_t \) antennas, an \gls{RIS} with \( N \) LC-based unit cells and $K$ single-antenna \glspl{MU}. The \glspl{MU} are served in a \gls{TDMA} scheme\footnote{In this paper, we do not address the time required for \gls{LC}-\gls{RIS} configuration \cite{delbari2025fast}; instead, our focus is solely on the loss-related trade-off.}. The received signal at $k$th \gls{MU} in their allocated time slot is given by:
\begin{equation}
    y_k = \left( \mathbf{h}_{d,k}^\Herm + \mathbf{h}_{r,k}^\mathsf{H} \bGamma_k \mathbf{H}_t \right) \mathbf{x}_k + n_k, \forall k=1, \cdots, K,
\end{equation}
where $\bx_k\in\Cset^{N_t}$ is the transmit signal vector for the $k$th \gls{MU}, $y_k\in\Cset$ is the received signal vector at the $k$th \gls{MU}, and $n_k\in\Cset$ represents the \gls{AWGN} at the $k$th \gls{MU}, i.e., $n_k\sim\sCN(0,\sigma_n^2),\,\forall k$, where $\sigma_n^2$ is the noise power. Assuming linear beamforming, the transmit vector $\bx_k$ can be written as $\bx_k=\bq_k s$, where $\bq_k\in\Cset^{N_t}$ is the beamforming vector on the \gls{BS}  for $k$th \gls{MU} and $s\in\Cset$ is the data symbol. Assuming $\Ex\{|s|^2\}=1$, the beamformer satisfies the transmit power constraint $\|\bq_k\|^2\leq P_k,\,\forall k$ with $P_k$ denoting the maximum required transmit power for $k$th \gls{MU}. Moreover,  $\bh_{d,k}^\Herm\in\Cset^{N_t}, \bH_t\in\Cset^{N\times N_t}$, and $\bh_{r,k}^\Herm\in\Cset^{N}$ denote the \gls{BS}-\gls{MU}, \gls{BS}-\gls{RIS}, and \gls{RIS}-\gls{MU} channel matrices, respectively.
The RIS reflection matrix is \( \bGamma_k = \diag([\bar{\bGamma}_k]_n e^{j[\bomega_k]_n}) \in \mathbb{C}^{N \times N} \), where $[\bomega_k]_n$ is $n$th element phase shifter for $k$th \gls{MU} and \( [\bar{\bGamma}_k]_n \) denotes the amplitude of $n$th \gls{RIS} element for $k$th \gls{MU} and accounts for the loss. 

\subsection{Channel Model}
\label{sec: channel model}
For extremely large \glspl{RIS}, the distances between the \gls{RIS} and both the \gls{BS} and the \gls{MU} may fall within the \gls{NF} regime of the \gls{RIS} \cite{delbari2024nearfield}. Consequently, an \gls{NF} channel model is adopted. Moreover, \glspl{RIS} are typically installed at elevated heights, ensuring \gls{LOS} connectivity between the \gls{RIS} and both the \gls{BS} and the \glspl{MU}. The use of high-frequency bands further strengthens the dominance of \gls{LOS} links over \gls{nLOS} links. As a result, the channels are modeled using Rician fading with a high $K$-factor, representing the strong contribution of \gls{LOS} components relative to \gls{nLOS} components.

For clarity, we present the model for a general $\bH \in \Cset^{N_\rx \times N_\tx}$, where $N_\tx$ and $N_\rx$ denote the number of transmit and receive antennas, respectively. Based on the aforementioned considerations, the channel can be approximated as $\bH \approx \bH^{\mathrm{LOS}}$, where
\begin{equation}
	[\bH^\LOS]_{m,n} = \, c_0\e^{\jj\kk\|\bu_{\rx,m}-\bu_{\tx,n}\|},\label{Eq:LoSnear}
\end{equation}
where $\bH^{\LOS}$ denotes the \gls{LOS} \gls{NF} channel matrix, $c_0$ represents the channel amplitude of the LOS path, and $\bu_{\tx,n}$ and $\bu_{\rx,m}$ are the locations of the $n$th \gls{Tx} antenna and the $m$th \gls{Rx} antenna, respectively. Furthermore, $\kk = 2\pi/\lambda$ is the wave number, where $\lambda$ is the wavelength. In addition, we assume \gls{BS}–\glspl{MU} direct paths are blocked, i.e., $\bh_{d,k}^\Herm\approx\bzero,\,\forall k$. The mentioned channel model can be applied to \( \mathbf{H}_t \) and \( \mathbf{h}_{r,k} \), \( \forall k \in \{1, \cdots, K\} \).

\begin{figure}[t]
    \centering
    \includegraphics[width=0.5\textwidth]{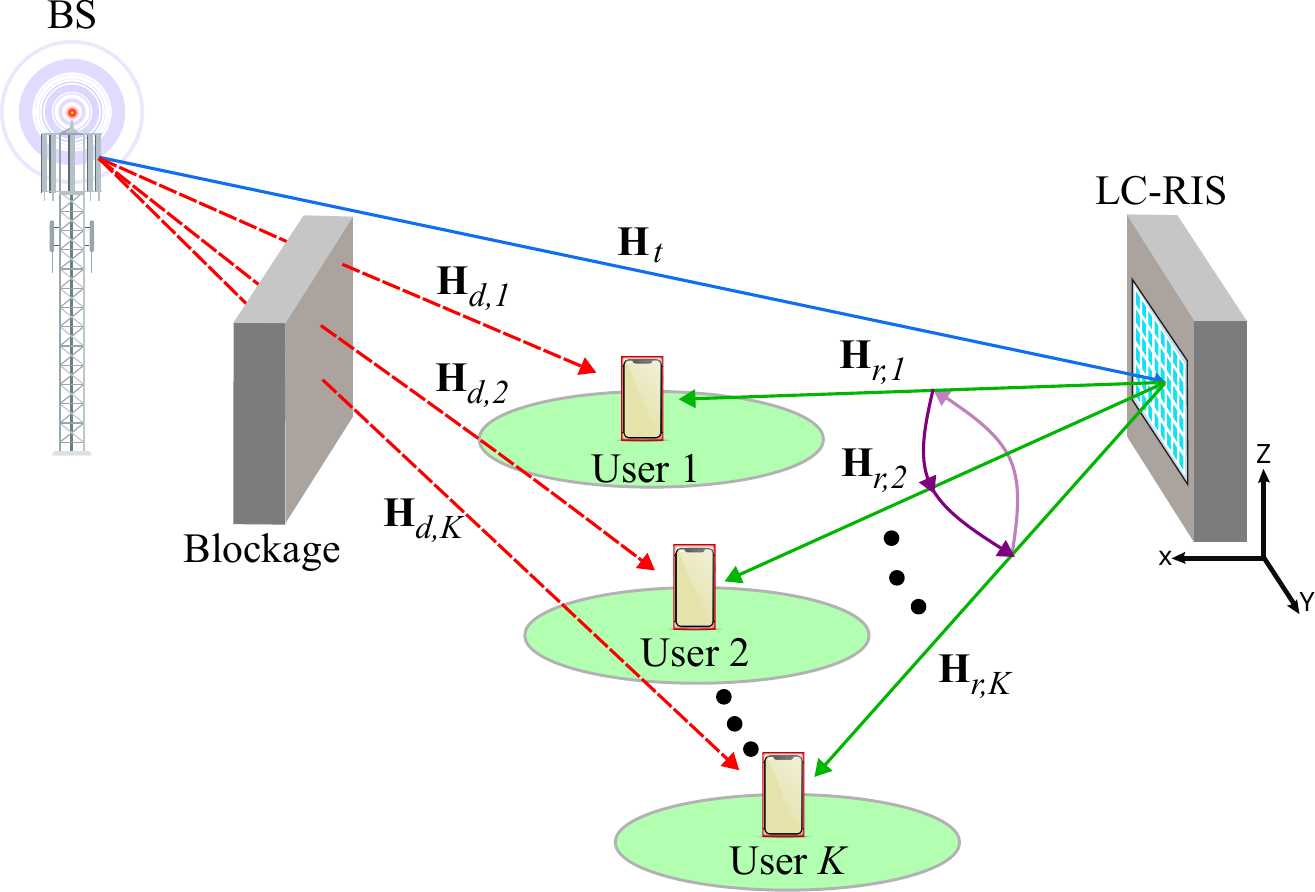}
    \caption{A schematic of a wireless channel model where a \gls{BS} serves multiple users in different locations via an \gls{RIS}.}
    \label{fig:system model}
    \vspace{-0.5cm}
\end{figure}

\subsection{Data Rate}
The data rate for $k$th user is defined as follows\footnote{Here, we did not include the interference as we use \gls{TDMA} scheme \cite{delbari2024fast}.}:

\begin{equation}
    R_k = \log(1 + \text{SNR}_k),
\end{equation}
\begin{equation}
    \text{SNR}_k = \frac{|\left(\mathbf{h}_k^{\text{eff}}\right)^\Herm \bq_k|^2}{\sigma_n^2}, \quad \text{with } (\mathbf{h}_k^{\text{eff}})^\Herm = \mathbf{h}_{r,k}^\Herm \bGamma_k \mathbf{H}_t.
\end{equation}

To enhance the data rate, we jointly and iteratively optimize $\bGamma_k$ and $\bq_k$. In contrast to most existing works that assume full \gls{CSI}, we consider only the dominant \gls{LOS} link when optimizing the phase shifts of the \gls{LC}-\gls{RIS} in \gls{mmWave} systems. Moreover, to reduce the overhead associated with frequent optimization, we consider area sets, $\Pset_k,\,\forall k$, which represent the possible locations of the $k$th \gls{MU}, instead of considering them as single fixed points \cite{delbari2024far}.

\section{Loss Model in LC-RIS Phase Shifter}
\label{sec: Loss Model in LC-RIS Phase shifter}


The maximum achievable phase-shift range of an \gls{LC} unit cell is governed by multiple physical parameters, which is described by:

\begin{equation}
\label{eq: omega and l}
    \Delta \omega_{\text{max}} = 2\pi l \Delta n  \frac{f}{c},
\end{equation}
where $\Delta n$, \( l \), \( f \), and \( c \) denote the birefringence, physical length of the phase shifter, the operating frequency, and the speed of light in vacuum, respectively. In an ideal case, $\Delta \omega_{\text{max}}=2\pi$ when $l=l_r$ where $l_r$ is the reference \gls{LC} phase shifter length. Therefore, for different lengths, we can derive the maximum achievable phase shift as follows:
\begin{equation}
    \Delta \omega_{\text{max}}=2\pi\frac{l}{l_r}.
    \label{eq: l_r}
\end{equation}

An important metric for evaluating LC-based phase shifter performance is the \gls{FOM} \cite{neuder2023compact,neuder2024architecture}, defined as:
\begin{equation}
    \FoM = \frac{\Delta \omega_{\max}}{|\Omega|^2_\dB},
\end{equation}
where \( |\Omega|^2_\dB \) is the maximum insertion loss at a specific operating frequency in dB. Since the $\FoM$ is constant at a given frequency, a shorter LC phase shifter length \( l \) leads to a decrease of the maximum insertion loss $|\Omega|^2_\dB$. Accordingly, the insertion loss \gls{w.r.t.} \( l \) is given by:
\begin{equation}
    |\Omega|^2_\dB = \Big(\frac{2\pi}{\FoM}\Big)\frac{l}{l_r}~\text{dB}.
    \label{eq:Pl}
\end{equation}

Eqs. \eqref{eq: l_r} and \eqref{eq:Pl} suggest a fundamental trade-off such that decreasing $l$ reduces the insertion loss but at the same time reduces $\Delta\omega_{\max}$, which negatively impacts the performance.

\section{LC-RIS Loss Trade-off}
Note that both $\Delta\omega_{\max}$ and $|\Omega|^2_\dB$ impact the required transmit power at the \gls{BS} to meet a certain \gls{QoS} requirement. Therefore, to investigate this trade-off, we study how to configure the BS beamformer and the LC-RIS phase shifts such that the transmit power is minimized while a given \gls{SNR} is met. Specifically, we assume $|[\bar{\bGamma}_k]_n|^2=1/|\Omega_k|^2,\,\forall k, n$, where $|\Omega_k|^2_\dB=10\log_{10}(|\Omega_k|^2)$ and $\Delta\omega_{\max}=\omega_{\max}$. Under this assumption, the problem becomes:
\begin{subequations}
\label{eq:optimization 1}
\begin{align}
    \text {P1:}\quad&~\underset{\bomega_k,\bq_k,l_k}{\min}~P_k
    \\&~\text {s.t.} ~~ \text{ C1: }\SNR_k\geq \SNR_\thr,\, \forall \bp_k\in\Pset_k, \forall k,
    \\&\quad\hphantom {\text {s.t.} } \text{ C2: } 0\leq [\bomega_k]_n < \omega_{\tmax,k}, \forall k,n,
    \\&\quad\hphantom {\text {s.t.} } \text{ C3: } \|\bq_k\|_2^2\leq P_k,\,\forall k,
\end{align}
\end{subequations}
  where $\bp_k$ is the location of the $k$th \gls{MU}\footnote{In a dynamic scenario, the $k$th \gls{MU} can be localized first, after which the proposed algorithm selects the optimal beamformer, \gls{RIS} phase-shift configuration, and \gls{LC}-\gls{RIS} phase-shift length.}, C1 is the $k$th \gls{MU} \gls{SNR} constraint, C2 is a constraint over phase shifts which is related to the length of the \gls{LC} phase shifter, and C3 is the transmitted beamformer constraint. $\SNR_\thr$ is the minimum required \gls{SNR} that must be satisfied for all the \glspl{MU} and $P_k=|\Omega_k|^2P$ where $P$ is the required power regardless of the loss to meet required \gls{SNR} (i.e., in a lossless scenario). Problem P1 is non-convex due to the non-convexity of C1 in $\bomega_k$. In addition, for all \glspl{MU}, $l_k$ is coupled with $\bomega_k$ and $\bq_k$ in C2 and C3 constraints, respectively. This makes it more challenging to find a global solution. To address this, we decompose P1 into two subproblems and iteratively minimize the cost function for each \gls{MU} using \gls{AO}, i.e., solving the problem separately for the $k$th \gls{MU}.
\subsection{Beamformer Design}
In this step, we fix $\bomega_k,\forall k$ and optimize only $\bq_k$ in problem P1. Under this assumption, problem P1 reduces to
\begin{subequations}
	\label{eq:optimization 2}
	\begin{align}
		\text {P2:}\quad&~\underset{\bq_k}{\min}~P_k
		\\&~\text {s.t.} ~~ \text{ C1 and C3}.
	\end{align}
\end{subequations}
For a given phase-shift vector $\bomega_k$, it is known that \gls{MRT} is the optimal transmit beamformer \cite{tse2005fundamentals}, i.e.,
\begin{equation}
\label{eq: bq_k}
\bq_k=\frac{\bh_k^\eff}{\|\bh_k^\eff\|}\sqrt{P_k},
\end{equation}
where $P_k=\frac{\sigma_n^2}{\|\bh_k^\eff\|^2}\SNR_\thr$.

\subsection{RIS Phase-shift Configuration}
When the beamformer is fixed, Problem~P1 becomes equivalent to maximizing $\|\bh_k^\eff\|,\forall k$ or equivalently maximizing $\SNR_k(\bp_k),\,\forall k,\,\forall \bp_k\in\Pset_k$. To formalize this, we introduce an auxiliary variable $\alpha$ and reformulate the problem as
\begin{subequations}
	\label{eq:optimization 3}
	\begin{align}
		\text {P3:}\quad&~\underset{\bomega_k,l_k}{\max}~\alpha,
        \\&~\text {s.t.} ~~ \widehat{\text{C1}}: \SNR_k\geq\alpha,\,\forall k,\,\forall \bp_k\in\Pset_k, \text{C2}.
	\end{align}
\end{subequations}
We address Problem~P3 for fixed values of $l_k$ corresponding to $\omega_{\max,k} \in (0, 2\pi)$. Specifically, we first fix $\omega_{\max,k}$ according to a given $l_k$ determined by \eqref{eq: l_r} for $k$th \gls{MU}.
Once $\omega_{\max,k}$ is fixed, the problem can be solved following the approach in~\cite[Algorithm~1]{Delbari2024temperature}. To this end, we express $\SNR_k$ in terms of the \gls{RIS} phase-shift vector $\bs_k\defeq[\e^{\jj[\bomega_k]_1}, \cdots, \e^{\jj[\bomega_k]_N}]^\Trans$. With this definition, we have
\begin{subequations}
    \label{eq: SNR in term of s}
    \begin{align}
        \SNR_k=&\bs_k^\Herm\bA_k\bs_k,
    \end{align}
\end{subequations}
where $\bA_k=\frac{\diag(\bh_{r,k}^\Herm)\bH_t\bq_k\bq_k^\Herm\bH_t^\Herm\diag(\bh_{r,k})}{\sigma_n^2},\,\forall k$. We further define $\bS_k=\bs_k\bs_k^\Herm,\forall k$, which leads to the equivalent formulation
\begin{subequations}
\label{eq:optimization 4}
\begin{align}
    \text {P4:}&~\underset{\bS_k}{\max}~\alpha
    \\&~\text {s.t.} ~~\widehat{\widehat{\text{C1}}}:\tr(\bA_k\bS_k)\geq\alpha, \forall \bp_k\in\Pset_k, \text{C2},
    \\&\quad\hphantom {\text {s.t.} }\text{C3: } \bS_k\succeq 0,\text{C4: } \rank(\bS_k)=1, \text{C5: }\diag(\bS_k)=\bone_N.
\end{align}
\end{subequations}
Problem~P4 remains non-convex due to (i) the non-convexity of C2 in $\bS_k$ and (ii) the rank-one constraint C4. To address the rank-one constraint, we employ the penalty method from \cite{Yu2020power}. To handle the non-convexity of C2 in $\bS_k$, we apply the reformulation in \cite[Lemma~2 and Lemma~3]{Delbari2024temperature}, yielding, $\widehat{\text{C2}}:,\,\forall n,k$
\begin{align}
    &\real(\zeta\sum_{i=1}^N [\bS_k]_{n,i})\!+\!\tan(\frac{\omega_{\max}}{2})\imag(\zeta\sum_{i=1}^N [\bS_k]_{n,i})\leq1, \!\!\!\!\!\!\!\!\!&\omega_{\max}>\pi,\nonumber\\
    &2\cos(\frac{\omega_{\max}}{2})\imag(\frac{\zeta}{2}\sum_{i=1}^N [\bS_k]_{n,i})\leq\imag(\zeta\sum_{i=1}^N [\bS_k]_{n,i}), &\omega_{\max}\leq\pi,
\end{align}
where $\zeta\defeq\frac{\jj\omega_{\max}}{N(1-\e^{-\jj\omega_{\max}})}$. With this transformation, Problem P4 can be reformulated as
\begin{subequations}
\label{eq:optimization 5}
\begin{align}
    \text {P5:}&~\underset{\bS_k}{\max}~\alpha-\eta^{(i)}\Big(\|\bS_k\|_*-\|\bS_k^{(i)}\|_2-\tr\big(\blambda_{\max}(\bS_k^{(i)})\nonumber
    \\&\quad\quad\quad\times\blambda_{\max}^\Herm(\bS_k^{(i)})(\bS_k-\bS_k^{(i)})\big)\Big)
    \\&~\text {s.t.} ~~\widehat{\widehat{\text{C1}}},\widehat{\text{C2}}, \text{C3}, \text{C5},
\end{align}
\end{subequations}
where $\eta^{(i)}$ is the penalty factor at iteration $i$ which increases gradually. By selecting a sufficiently large $\eta$, problems P4 and P5 become equivalent.

\subsection{Algorithm and Complexity Analysis}
The proposed algorithm is summarized in Algorithm \ref{alg:cap}. At each iteration, the most computational step is the computation of the nuclear norm in line 6, which has a complexity of $\bigO(N^3)$. The number of different constraints generated by $\widehat{\widehat{\text{C1}}}$ is the bottleneck and proportional to $|\Pset|=\underset{k}{\max} |\Pset_k|$. Thus, the complexity of the Algorithm \ref{alg:cap} in total is $\bigO(I_{\max}K|\Pset||\Wset|N^3)$, where $|\Wset|$ is the number of samples in $\omega_{\max}$ in line 3.

\begin{algorithm}[t]
\caption{Proposed Algorithm for Problem P1}\label{alg:cap}
\begin{algorithmic}[1]
\STATE \textbf{Initialize: $\bp_k,\, I_{\max},\, \epsilon$.}
\FOR{$k=1, \cdots, K$}
\FOR{$\omega_{\max,k}=0, \cdots, 2\pi$}
\STATE Set $i=1$ and $\bs_k^{(0)}\!=\!\e^{\jj\omega_{\max,k}\times\mathrm{rand}(N)},\bS_k^{(0)}=\bs_k^{(0)}{\bs_k^{(0)}}^\Herm$.
\WHILE{$\|\bS_k^{(i)}-\bS_k^{(i-1)}\|_F^2\geq\epsilon$ and $i\leq I_{\tmax}$}
    \STATE Solve convex P5 for given $\bS_k^{(i-1)}$, and store the intermediate solution $\bS_k^{(i)}$.
    \STATE Set $i = i + 1$ and update $\eta^{(i)} =5\eta^{(i-1)}$.
\ENDWHILE
\ENDFOR
\STATE $[\bS_k^*,\omega_{\max,k}^*]\defeq\underset{\bS_k,\omega_{\max,k}}{\text{argmax}}~\alpha,$ according to P3.
\STATE $\bq_k$ is calculated with \eqref{eq: bq_k}.
\ENDFOR
\end{algorithmic}
\end{algorithm} 

\section{Performance Evaluation}
By solving problem P5, we show the trade-off between loss and total transmit power under several fixed $\omega_{\max}$ values. This trade-off highlights a key system design insight: as the length of the phase shifter decreases, the maximum achievable phase shift also decreases. While this reduces the loss in \gls{LC} cells, the \gls{BS} may have to transmit at a higher power to maintain the same level of data rate performance.

\subsection{Simulation Setup} 
In our simulation, the \gls{RIS} center is the origin of the Cartesian coordinate system, i.e., $[0,0,0]$~m. We assume there are four \glspl{MU} in $[10, y, -5]$~m where $y=0, 1, 2, 5$ with radius $R$ meter. Each \gls{MU}'s region is sampled with a spatial resolution of $0.5~\text{m}$. The BS comprises a $4 \times 4 = 16$ \gls{UPA} located at $[20,0,10]$~m. The RIS is a two-dimensional \gls{UPA} consisting of $N_y \times N_z = 10 \times 10$ elements aligned to the y and z axes, respectively. The element spacing for both the BS and RIS is half of the wavelength. The noise variance is computed as $\sigma_n^2 = W N_0 N_f$, where $N_0 = -174$~dBm/Hz, $W = 20$~MHz, and $N_f = 6$~dB. We assume a 28~GHz carrier frequency and adopt the path loss model $\rho(d_0/d)^\sigma$, where $\rho = -61$~dB at $d_0 = 1$~m. Moreover, we set $\sigma = (2,2,2)$ and $K = (0,10,10)$ for the BS-MU, BS-RIS, and RIS-MU channels, respectively. This scenario illustrates the trade-off between SNR and $P_k$, as varying the phase shifter length directly influences the RIS design and its efficiency in data rate.
\subsection{Simulation Results} 
\begin{figure}
    \centering
    \includegraphics[width=0.5\textwidth]{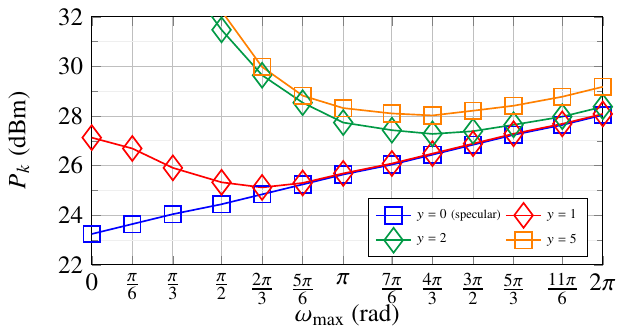}
    \caption{Impact of maximum differential phase-shift range (insertion loss) on required total transmit power for four different \glspl{MU} with $\SNR_\thr=10$~dB.}
    \vspace{-0.5 cm}
    \label{fig:sr_vs_loss}
\end{figure}
As shown in Fig.~\ref{fig:sr_vs_loss}, we evaluate the impact of insertion loss on the required transmit power when the \gls{RIS} must satisfy a \gls{QoS} requirement of $\SNR_\thr = 10$~dB for four \glspl{MU} located at different positions. Based on the data of the DGS-IMSL, the $\mathrm{FoM}$ corresponds to $75^\circ/\mathrm{dB}$ at the considered frequency \cite{neuder2023compact}. For this simulation, we set the radius to its smallest possible value (focus point). Since \gls{MU} 1 lies in the specular reflection direction of the \gls{RIS}, a mirror-like behavior, achieved when all phase shifters are set to zero, yields the optimal configuration. In this case, the \gls{MU} requirement is satisfied without any additional phase shift, and increasing the phase-shift range only introduces more insertion loss without improving the \gls{SNR}. In contrast, other \glspl{MU} are not in the specular direction, so the optimal phase configuration for the \gls{RIS} is more complex. A broader phase-shift range is needed to satisfy the \gls{MU} requirement in this case. Specifically, as the \gls{MU}’s position deviates further from the specular angle, the necessary maximum phase shift, $\omega_{\max}$, increases correspondingly.

\begin{figure}
    \centering
    \includegraphics[width=0.5\textwidth]{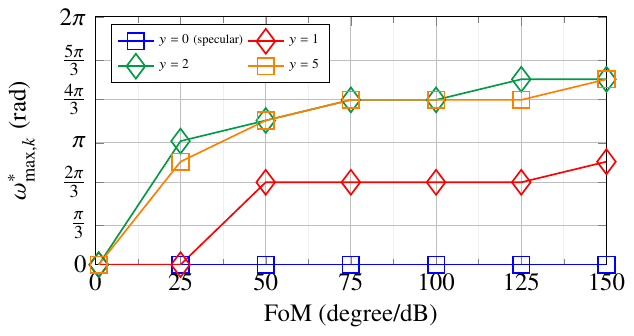}
    \caption{Impact of the \gls{FOM} value on the optimized $\omega_{\max}$.}
    \vspace{-0.5 cm}
    \label{fig:FOM}
\end{figure}

Fig.~\ref{fig:FOM} depicts the optimal maximum differential phase-shift range $\omega_{\max,k}^\ast$ as a function of the LC phase shifter's \gls{FOM} for different MU positions. The \gls{FOM}, expressed in degrees per dB, quantifies the trade-off between achievable phase-shift range and insertion loss, as defined in Section~\ref{sec: Loss Model in LC-RIS Phase shifter}. A larger \gls{FOM} indicates reduced insertion loss for a given phase-shifter length, enabling a broader feasible phase-shift range. For MUs located near the specular reflection direction (e.g., first \gls{MU} with $y=0$), the required $\omega_{\max,k}^\ast$ is small and remains almost constant over a wide range of \gls{FOM} values, since only minor phase adjustments are necessary to satisfy the SNR constraint. In contrast, MUs positioned away from the specular reflection path (e.g., MU~$y>0$) exhibit a significant increase in $\omega_{\max,k}^\ast$ with \gls{FOM}, reflecting the need for larger phase compensation.

\begin{figure}
    \centering
    \includegraphics[width=0.5\textwidth]{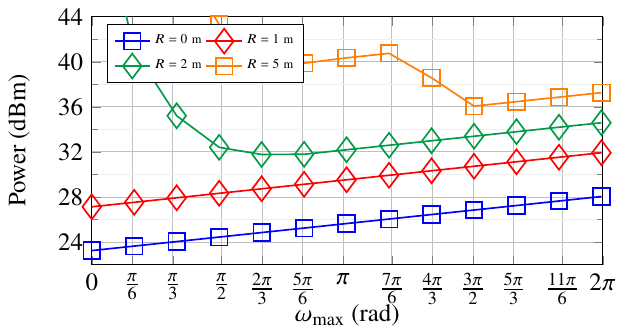}
    \caption{Impact of four different coverage sizes on the optimal maximum phase shift ($\omega_{\max}$).}
    \label{fig:area}
    \vspace{-0.5 cm}
\end{figure}

Fig.~\ref{fig:area} shows the required transmit power as a function of $\omega_{\max}$ for four different coverage area sizes when the \gls{RIS} serves the first \gls{MU}. As the radius of the coverage area ($R = |\Pset|$) increases, both the required transmit power and the optimal value of $\omega_{\max}$ increase.

\section{Conclusion and Future Work}
This study investigated a key trade-off in \gls{LC}-based \gls{RIS} design. While a shorter \gls{LC} length reduces insertion loss, it also restricts phase control, thereby affecting system performance. To isolate and highlight this trade-off, we considered a simplified scenario focused on satisfying the \gls{QoS} requirements of multiple users. However, this trade-off remains relevant and can be further explored in more complex and practical applications as well.

\bibliographystyle{IEEEtran}
\bibliography{References}

\begin{thebibliography}{10}
\providecommand{\url}[1]{#1}
\csname url@samestyle\endcsname
\providecommand{\newblock}{\relax}
\providecommand{\bibinfo}[2]{#2}
\providecommand{\BIBentrySTDinterwordspacing}{\spaceskip=0pt\relax}
\providecommand{\BIBentryALTinterwordstretchfactor}{4}
\providecommand{\BIBentryALTinterwordspacing}{\spaceskip=\fontdimen2\font plus
\BIBentryALTinterwordstretchfactor\fontdimen3\font minus \fontdimen4\font\relax}
\providecommand{\BIBforeignlanguage}[2]{{%
\expandafter\ifx\csname l@#1\endcsname\relax
\typeout{** WARNING: IEEEtran.bst: No hyphenation pattern has been}%
\typeout{** loaded for the language `#1'. Using the pattern for}%
\typeout{** the default language instead.}%
\else
\language=\csname l@#1\endcsname
\fi
#2}}
\providecommand{\BIBdecl}{\relax}
\BIBdecl

\bibitem{jimenez2023reconfigurable}
A.~Jim{\'e}nez-S{\'a}ez \emph{et~al.}, ``Reconfigurable intelligent surfaces with liquid crystal technology: A hardware design and communication perspective,'' \emph{preprint arXiv:2308.03065}, 2023.

\bibitem{jakoby2020microwave}
R.~Jakoby, A.~Gaebler, and C.~Weickhmann, ``Microwave liquid crystal enabling technology for electronically steerable antennas in satcom and {5G} millimeter-wave systems,'' \emph{Crystals}, vol.~10, no.~6, p. 514, 2020.

\bibitem{neuder2024architecture}
R.~Neuder, M.~Sp{\"a}th, M.~Sch{\"u}{\ss}ler, and A.~Jim{\'e}nez-S{\'a}ez, ``Architecture for sub-100 ms liquid crystal reconfigurable intelligent surface based on defected delay lines,'' \emph{Commun. Eng.}, vol.~3, no.~1, p.~70, 2024.

\bibitem{Huang2019}
C.~Huang \emph{et~al.}, ``Reconfigurable intelligent surfaces for energy efficiency in wireless communication,'' \emph{IEEE Trans. Wireless Commun.}, vol.~18, no.~8, pp. 4157--4170, 2019.

\bibitem{Tang2021}
W.~Tang \emph{et~al.}, ``Wireless communications with reconfigurable intelligent surface: Path loss modeling and experimental measurement,'' \emph{IEEE Trans. Wireless Commun.}, vol.~20, no.~1, pp. 421--439, 2021.

\bibitem{Singh2023}
I.~Singh, P.~J. Smith, and P.~A. Dmochowski, ``Phase dependent loss analysis for {RIS} systems,'' in \emph{Proc. IEEE Wireless Commun. and Netw. Conf. (WCNC)}, 2023, pp. 1--6.

\bibitem{Ndjiongue2021}
A.~R. Ndjiongue \emph{et~al.}, ``Design of a power amplifying-ris for free-space optical communication systems,'' \emph{IEEE Wireless Communications}, vol.~28, no.~6, pp. 152--159, 2021.

\bibitem{delbari2025fast}
M.~Delbari, R.~Neuder, A.~Jim{\'e}nez-S{\'a}ez, A.~Asadi, and V.~Jamali, ``Fast reconfiguration of {LC-RISs}: Modeling and algorithm design,'' \emph{arXiv preprint arXiv:2504.08352}, 2025.

\bibitem{delbari2024nearfield}
M.~Delbari, G.~C. Alexandropoulos, R.~Schober, H.~V. Poor, and V.~Jamali, ``Near-field multipath {MIMO} channel model for imperfect surface reflection,'' in \emph{IEEE Globecom Workshops (GC Wkshps)}, 2024.

\bibitem{delbari2024fast}
M.~Delbari \emph{et~al.}, ``Fast transition-aware reconfiguration of liquid crystal-based {RIS}s,'' in \emph{Proc. IEEE ICC Workshops}, 2024, pp. 214--219.

\bibitem{delbari2024far}
M.~Delbari, G.~C. Alexandropoulos, R.~Schober, and V.~Jamali, ``{Far- versus Near-Field RIS Modeling and Beam Design},'' in \emph{Reconfigurable Metasurfaces for Wireless Communications: Architectures, Modeling, and Optimization}.\hskip 1em plus 0.5em minus 0.4em\relax Springer, 2024, arXiv preprint: \url{https://arxiv.org/pdf/2401.08237}.

\bibitem{neuder2023compact}
R.~Neuder \emph{et~al.}, ``Compact liquid crystal-based defective ground structure phase shifter for reconfigurable intelligent surfaces,'' in \emph{European Conf. Antennas and Propag. (EuCAP)}, 2023, pp. 1--5.

\bibitem{tse2005fundamentals}
D.~Tse, ``Fundamentals of wireless communication,'' \emph{Cambridge University Press google schola}, vol.~2, pp. 614--624, 2005.

\bibitem{Delbari2024temperature}
M.~Delbari \emph{et~al.}, ``Temperature-aware phase-shift design of {LC-RIS} for secure communication,'' in \emph{Proc. IEEE ICC}, 2025, arXiv:2411.12342.

\bibitem{Yu2020power}
X.~Yu \emph{et~al.}, ``Power-efficient resource allocation for multiuser miso systems via intelligent reflecting surfaces,'' in \emph{IEEE Global Commun. Conf. (GLOBECOM)}, 2020, pp. 1--6.

\end{thebibliography}

\end{document}